\documentclass[manuscript,screen, nonacm]{acmart}
\usepackage{multirow}
\usepackage{xcolor}  

\AtBeginDocument{%
  \providecommand\BibTeX{{%
    \normalfont B\kern-0.5em{\scshape i\kern-0.25em b}\kern-0.8em\TeX}}}

\copyrightyear{2023}
\acmYear{2023}
\setcopyright{acmlicensed}\acmConference[AutomotiveUI '23]{15th
International Conference on Automotive User Interfaces and Interactive
Vehicular Applications}{September 18--22, 2023}{Ingolstadt, Germany}
\acmBooktitle{15th International Conference on Automotive User Interfaces
and Interactive Vehicular Applications (AutomotiveUI '23), September 18--22,
2023, Ingolstadt, Germany}
\acmPrice{15.00}
\acmDOI{10.1145/3580585.3607167}
\acmISBN{979-8-4007-0105-4/23/09}

\begin{document}

\title[eHMI Scalability Issues]{Scoping Out the Scalability Issues of Autonomous Vehicle-Pedestrian Interaction}


\author{Tram Thi Minh Tran}
\email{tram.tran@sydney.edu.au}
\orcid{0000-0002-4958-2465}
\affiliation{Design Lab, Sydney School of Architecture, Design and Planning,
  \institution{The University of Sydney}
  \city{Sydney}
  \state{NSW}
  \country{Australia}
}

\author{Callum Parker}
\email{callum.parker@sydney.edu.au}
\orcid{0000-0002-2173-9213}
\affiliation{Design Lab, Sydney School of Architecture, Design and Planning,
  \institution{The University of Sydney}
  \city{Sydney}
  \state{NSW}
  \country{Australia}
}

\author{Martin Tomitsch}
\email{martin.tomitsch@sydney.edu.au}
\orcid{0000-0003-1998-2975}
\affiliation{Design Lab, Sydney School of Architecture, Design and Planning,
  \institution{The University of Sydney}
  \city{Sydney}
  \state{NSW}
  \country{Australia}
}

\renewcommand{\shortauthors}{Tran et al.}

\begin{abstract}
Autonomous vehicles (AVs) may use external interfaces, such as LED light bands, to communicate with pedestrians safely and intuitively. While previous research has demonstrated the effectiveness of these interfaces in simple traffic scenarios involving one pedestrian and one vehicle, their performance in more complex scenarios with multiple road users remains unclear. The scalability of AV external communication has therefore attracted increasing attention, prompting the need for further investigation. This scoping review synthesises information from 54 papers to identify seven key scalability issues in multi-vehicle and multi-pedestrian environments, with Clarity of Recipients, Information Overload, and Multi-Lane Safety emerging as the most pressing concerns. To guide future research in scalable AV-pedestrian interactions, we propose high-level design directions focused on three communication loci: vehicle, infrastructure, and pedestrian. Our work contributes the groundwork and a roadmap for designing simplified, coordinated, and targeted external AV communication, ultimately improving safety and efficiency in complex traffic scenarios.

\end{abstract}




\maketitle

\section{Introduction}

Autonomous vehicles (AVs) are often hailed as a safer alternative to human drivers because they can react faster and more accurately to hazards on the road~\cite{anderson2014autonomous, fagnant2015preparing}. However, before AVs can be safely integrated into our transportation system, certain roadblocks must be overcome, including the ability of AVs to communicate and interact seamlessly with vulnerable road users (VRUs), such as pedestrians~\cite{googlecar2016, rasouli2019autonomous}. The absence of interpersonal cues, such as drivers' hand signals and eye contact, is likely to cause uncertainty and misunderstanding for pedestrians in ambiguous situations. To address this challenge, researchers and the automotive industry are investigating the use of external Human-Machine Interfaces (eHMIs) as a way for AVs to communicate with pedestrians~\cite{rouchitsas2019external, dey2020taming, de2022external}. For example, the Mercedes-Benz F015 concept projects a pedestrian crossing onto the road to provide pedestrians with important information about the vehicle's intended actions~\cite{mercedes2015}.

The study of AV external communication has seen significant growth in recent years, leading to a plethora of eHMI design concepts that can be classified in various ways~\cite{dey2020taming}. These interfaces are often evaluated using scenario-based approaches, initially focusing on basic encounters between a single AV and a pedestrian~\cite{colley2020unveiling, tran2021review}. Preliminary evidence suggests that eHMIs may help to alleviate uncertainty~\cite{hollander2019investigating} and foster trust~\cite{faas2020longitudinal} in AVs. As such, the automotive research community within the field of HCI is increasingly exploring the scalability of eHMIs in more complex traffic scenarios involving multiple road users. 

Existing literature on AV external communication includes several notable works that delve into the scalability of eHMIs. In their literature analysis, \citet{colley2020unveiling} examined evaluation setups of 38 studies with regard to scalability factors (number of pedestrians, vehicles, lanes, and noise). \citet{colley2020design} proposed a design space that incorporated situational factors that affect the scalability of eHMIs, such as communication relationships (one-to-one, one-to-many, many-to-one, and many-to-many). \citet{dey2020taming} developed a classification taxonomy for eHMIs and highlighted scalability-related dimensions of a design concept, including communication strategy (mass or targeted), communication resolution (clarity of the intended recipients), and the number of users that can be addressed at a time. However, to date, no comprehensive review of the scalability issues that may arise has been conducted, limiting a thorough understanding of the community's collective challenges to inform future research efforts.

 In this study, we conducted a scoping review of AV external communication from 2014 - 2022 to identify scalability issues that have been discussed and/or empirically observed. These issues are categorised into two groups: those caused by the increase in the number of AVs (as well as mixed traffic\footnote{Mixed traffic refers to situations in which AVs must interact with human-driven vehicles on the same roads.}) and those caused by the increase in the number of pedestrians. Considering the most pressing issues, we pinpointed essential communication qualities for eHMIs to excel in complex traffic scenarios and propose high-level research directions focused on three communication loci: \textit{vehicle}, \textit{infrastructure}, and \textit{pedestrian}.

\textit{Contribution Statement:} This study contributes to the domain of AV external communication within HCI by offering a comprehensive review of the scalability issues of eHMIs in complex traffic scenarios. Our analysis illustrates the attention these issues have received within the automotive research community and introduces high-level approaches to address them. The study aims to inform the development of more effective and scalable eHMIs, ultimately contributing to the safe integration of AVs into the transport systems of future cities.

\section{Methods}

We conducted a scoping review, a preliminary assessment of the available literature on a particular topic, following the methodological framework proposed by Arksey and O'Malley~\cite{arksey2005scoping}.

\subsection{Data Sources and Search Strategies}
We retrieved relevant literature on AV external communication from four major databases: ACM Digital Library, IEEE Xplore, ScienceDirect, and Google Scholar. We performed a keyword search (within the title, abstract, and author keywords) using a query adapted from \citet{dey2020taming}: \textit{`autonomous OR automated OR self-driving OR driverless AND car OR vehicle AND pedestrian AND interface OR interaction OR communication'} (see Appendix A for the exact query syntax used for each database). We did not use \textit{`scalability'} as a keyword because not all publications explicitly used this term. We limited the search results to publications dated between 2014 and 2022, which corresponds to a period of accelerated development in eHMI concepts~\cite{dey2020taming}. The last search date was October 22, 2022. 

Our search query yielded 1305 entries (ACM = 233, IEEE Xplore = 461, ScienceDirect = 111, Google Scholar\footnote{The search on Google Scholar returned a large number of entries 19,100; therefore, we decided to conclude the search at page 50 after three consecutive pages of not finding any relevant entries.} = 500). After importing research results to a spreadsheet and removing 230 duplicates, \textbf{1075} publications remained to be screened. 

\subsection{Eligibility Criteria and Paper Selection}

We included papers that met the following criteria:
\begin{itemize}
    \item Language: The paper must be written in English, the primary language of the research team.
    \item Publication type: We considered full conference papers, late-breaking work (work in progress), journal articles, and technical reports, allowing for a comprehensive and timely understanding of the topic. 
    \item Relevance: The paper should focus on AV external communication. Furthermore, it must discuss and/or investigate the scalability of eHMIs in complex traffic scenarios involving multiple vehicles or pedestrians.
\end{itemize}

Our goal was to identify not only distinct arguments but also monitor the frequency of scalability discussions, leading us to include publications with similar viewpoints. We excluded papers that, despite featuring a multi-user setup, did not contain any information about the scalability of eHMIs. For example, the virtual environment in the study by~\citet{bockle2017sav2p} included a vehicle in the adjacent lane alongside the autonomous shuttle; however, the presence of the additional vehicle was not the main focus of the research, it had no textual description, and it did not demonstrate any reported impact on the findings.

The first author conducted the screening of the 1075 papers in two stages. In the initial stage, titles and abstracts were assessed for their relevance to AV external communication, resulting in 230 papers. The subsequent stage involved a thorough review of full-text papers to determine the presence of eHMI scalability information, resulting in 46 papers. Additional eligible papers were incorporated from the references of the aforementioned related works~\cite{colley2020unveiling, colley2020design, dey2020taming}, as well as from recently published papers not initially captured in the search. This brought the total number of publications to be analysed to \textbf{54}. A PRISMA (Preferred Reporting Items for Systematic Reviews and Meta-Analyses) flow diagram, which illustrates the inclusion and exclusion of papers throughout each stage of the review process, along with a list of included publications, can be found in Appendix B.

\subsection{Data Charting and Data Analysis}

The first author charted the following details from each included paper: author, year, title and eHMI scalability information (quoted in the authors' original words). For papers featuring empirical studies that examined eHMIs in complex traffic situations, additional data on eHMI design concept, eHMI placement (locus), positions of additional pedestrians, traffic type (mixed or all AVs), and evaluation method were also recorded.

During the familiarisation phase with the data, it was observed that the information related to eHMI scalability issues was largely well-defined, suggesting a lower likelihood of varying interpretations. For instance, a statement like \textit{`the increase in the cognitive load imposed on a pedestrian as the number of AVs with LED boards increases'}~\cite{robert2019future} denotes an issue of information overload. As a result, a single coder with subject matter expertise was engaged to ensure uniformity throughout the analysis process. The first author undertook a thematic analysis~\cite{braun2006using} of the collected scalability information, applying the affinity diagramming method. This analysis process employed a bottom-up approach, consolidating discussions around similar issues and resulting in the identification of seven distinct themes corresponding to seven scalability issues: \textit{Information Overload}, \textit{Multi-Lane Safety}, \textit{Audibility}, \textit{Clarity of Recipients}, \textit{Group Influence}, \textit{Visibility}, and \textit{Privacy}. Additionally, a top-down approach was utilised to categorise these issues into two groups: those that arise in scenarios involving multiple vehicles, and those surfacing in scenarios with multiple pedestrians.

To mitigate the risk of bias or oversight in this setup, weekly discussions were scheduled with the second author during the coding process. Upon identifying the final issues, they were presented, along with their associated scenarios and underlying causes, to the remaining authors to determine whether they agreed with these findings or had additional insights. Although the identified scalability issues remained unchanged after the validation process, several critical discussions emerged. Primarily, the top-down division may inadvertently simplify the intricate nature of the interactions at play. Indeed, some issues may not be strictly confined to either a multi-vehicle or a multi-pedestrian scenario; instances where they cross these boundaries exist. Consider \textit{Information Overload}, this issue is classified under multi-vehicle scenarios, but it could also occur in a multi-pedestrian situation. Imagine a single AV that must communicate with multiple pedestrians dispersed in various locations. Simultaneously processing these multiple signals could lead to information overload for the pedestrian. Despite this potential overlap, we maintain the division to highlight the context where the issue is most likely to surface, thus facilitating a more targeted approach to addressing these issues.

\section{Results}

\subsection {Scalability Issues}

In \autoref{tab:casesummary}, we categorised eHMI-related scalability issues into those relating to \textit{multiple vehicles} and those concerning \textit{multiple pedestrians}. We also highlighted empirical studies that examined eHMIs in multi-vehicle and/or multi-pedestrian scenarios. In \autoref{tab:studies}, we summarised the commonalities of these empirical studies.

\begin{table*}[t]
  \footnotesize
  \caption{Scalability issues of eHMIs in multi-vehicle and multi-pedestrian scenarios.}
  \label{tab:casesummary}
  \begin{tabular}{clp{7.2cm}}
    \toprule
    &\textbf{Scalability issues}&\textbf{Studies\textsuperscript{*}}\\
    \midrule
    \multirow{3}{*}{\includegraphics[scale=0.04]{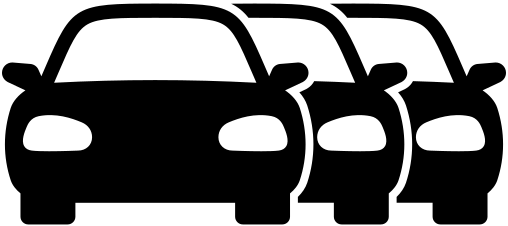}}
    &\textbf{(15) Information Overload} & 
    \textbf{\cite{hesenius2018don}
    \cite{hollander2022take}
    \cite{hollander2020save}
    \cite{rossi2022impact}
    \cite{tran2022designing}}
    \cite{dey2022investigating} 
    \cite{dey2020taming}
    \cite{hollander2019overtrust}
    \cite{mahadevan2018communicating}
    \cite{moore2019case}
    \cite{robert2019future}
    \cite{singer2020displaying} 
    \cite{verma2019pedestrians}
    \cite{verstegen2021commdisk}
    \cite{wang2020adaptability}
    \\
    &\textbf{(14) Multi-Lane Safety} &
    \textbf{\cite{colley2020towards}
    \cite{hollander2022take}
    \cite{mahadevan2019av}}
    \cite{andersson2017hello}
    \cite{colley2020evaluating}
    \cite{de2022external}
    \cite{dey2020taming}
    \cite{eisma2021external}
    \cite{habibovic2018communicating}
    \cite{hollander2019investigating}
    \cite{lagstrom2016avip}
    \cite{locken2019should}
    \cite{mok2022stopping}
    \cite{moore2019case}
    \\
    &\textbf{(3) Audibility} & 
    \textbf{\cite{colley2020towards}
    \cite{mahadevan2019av}}
    \cite{mahadevan2018communicating}
    \\
    \addlinespace[1em]
    \multirow{5}{*}{\includegraphics[scale=0.05]{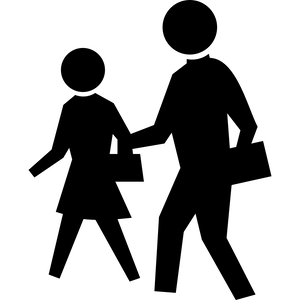}}
    &\textbf{(30) Clarity of Recipients} &
    \textbf{\cite{colley2021investigating}
    \cite{dey2021towards}
    \cite{dietrich2019projection}
    \cite{hoggenmueller2022designing}
    \cite{hollander2022take}
    \cite{hollander2020save}
    \cite{tran2022designing}
    \cite{wilbrink2021scaling}}
    \cite{ackermann2019experimental}
    \cite{ackermans2020effects}
    \cite{benderius2017best}
    \cite{de2022external}
    \cite{dey2020taming}
    \cite{dey2020distance}
    \cite{dey2018interface}
    \cite{eisma2021external}
    \cite{gui2022going}
    \cite{hensch2019effects}
    \cite{hollander2019investigating}
    \cite{kaleefathullah2022external}
    \cite{kass2020standardized}
    \cite{lagstrom2016avip}
    \cite{li2018cross}
    \cite{prattico2021comparing}
    \cite{robert2019future}
    \cite{singer2020displaying}
    \cite{stromberg2018designing}
    \cite{tabone2021vulnerable}
    \cite{tabone2021towards}
    \cite{verstegen2021commdisk}
    \\
    &\textbf{(5) Group Influence} &
    \textbf{\cite{colley2022effects}
    \cite{joisten2021communication}
    \cite{mahadevan2019av}
    \cite{chen2020autonomous}}
    \cite{lee2022learning}
    \\
    &\textbf{(5) Visibility} & 
    \textbf{\cite{troel2019ehmi}}
    \cite{asha2021co}
    \cite{ezzati2021interaction}
    \cite{risto2017human}
    \cite{verstegen2021commdisk}
    \\
    &\textbf{(1) Privacy} &
    \cite{lim2022ui} 
    \\
  \bottomrule
  \addlinespace
  \multicolumn{3}{l}{\textsuperscript{*}\textbf{Bold citations} refer to empirical studies that examined eHMIs in multi-vehicle and/or multi-pedestrian scenarios. }
  \end{tabular}
\end{table*}

\begin{table*}[t]
  \footnotesize
  \caption{Empirical studies examining eHMIs in multi-vehicle or multi-pedestrian scenarios.}
  \label{tab:studies}
  \begin{tabular}{lp{3.15cm}p{2.2cm}lll}
    \toprule
    \textbf{References} &\textbf{eHMI} &\textbf{Locus} & \textbf{Pedestrians\textsuperscript{*}} &\textbf{Vehicles}&\textbf{Environment\textsuperscript{**}}\\
    \midrule
    \citet{troel2019ehmi} & Lateral LED display & Vehicle & - & Multi AVs & Virtual reality (VR) \\
    \citet{rossi2022impact} & LED light band & Vehicle & - & Multi AVs & Test track\\
    \citet{hesenius2018don} & Wearable AR & Pedestrian & - & Multi AVs & Image \\
    \citet{tran2022designing} & Wearable AR & Pedestrian & - & Multi AVs & VR\\
    \citet{colley2020towards} & Auditory message & Vehicle & - & Multi AVs & VR\\
    \citet{hollander2022take} & Projection, Smart curbs & Vehicle, Infrastructure & Opposite side & Multi AVs & VR \\
    \citet{dietrich2019projection} & Projection, Signaling light  & Vehicle & Opposite side & - & VR \\
    \citet{colley2021investigating} & Texts on windshield & Vehicle & Opposite side & - & VR \\
    \citet{wilbrink2021scaling} & LED light band & Vehicle & Same, Opposite side & - & Video \\
    \citet{hollander2020save} & Smartphone & Pedestrian & Same side & Multi AVs & Video \\
    \citet{hoggenmueller2022designing} & LED light band & Vehicle & Same side & - & VR \\
    \citet{dey2021towards} & Contextual interfaces & Vehicle & Same side & - & VR\\
    \citet{colley2022effects} & LED light band & Vehicle & Group & - & VR \\
    \citet{joisten2021communication} & Walking man, Smiling face & Vehicle & Group & - & VR \\
    \citet{chen2020autonomous} & LED light band & Vehicle & Group & Mixed traffic & Public roads \\
    \citet{mahadevan2019av}
    &Mixed (LED light, physical hand, haptic cue, auditory message) & Vehicle, Infrastructure, Pedestrian & Group & Mixed traffic & VR\\
    \bottomrule
    \addlinespace
    \multicolumn{6}{l}{\textsuperscript{*}Additional pedestrian(s) in the scenario: forming a group with the participant or standing at a distance (same side or opposite side). }\\
    \multicolumn{6}{l}{\textsuperscript{**}Evaluation environment}\\
  \end{tabular}
\end{table*}

\subsubsection{Information Overload}

This concern has been raised as early as initial investigations into eHMIs. The term generally refers to a situation where the amount of information provided exceeds a person's cognitive processing capacity, resulting in poor decision quality~\cite{gross1964managing, eppler2004concept}. \citet{moore2019case} were among the scholars arguing that the eHMIs might transform street-crossing into an \textit{`analytical process'} and suggested the use of implicit motion cues for routine interactions. In classifying 70 eHMI concepts proposed by industry and academia, \citet{dey2020taming} also found that many designs utilised multiple modalities and displays (i.e., communication devices) and conveyed numerous messages. The authors thus emphasised the crucial need for creating a balance between leveraging redundancy (to foster interpretation and accuracy of use) and avoiding information overload. \citet{mahadevan2018communicating} confirmed this issue, where the study participants responded unfavourably to a mixed interface consisting of LED lights on the street, a printed hand mounted on the vehicle, and an auditory alert from a smartphone. The interface was deemed confusing and time-consuming as pedestrians had to wait for all go-ahead signals to be activated. Nonetheless, reducing the number of stimuli may not necessarily result in a reduced cognitive load. For example, participants in a study by \citet{dey2022investigating} preferred to have eHMIs communicate an AV's intent at all times rather than only during interactions (e.g., giving way).

A proliferation of AVs on urban streets is expected to exacerbate the issue of information overload in several ways. First, without a standardised universal set of symbols, manufacturers may adopt different interface designs to communicate a vehicle's awareness and intents \cite{verma2019pedestrians}. Second, a dynamic traffic scenario may involve AVs with various levels of automation (e.g., partially and fully automated) ~\cite{hollander2019overtrust, hollander2020save} and in different states of operation (e.g., cruising and stopping)~\cite{dey2022investigating}. The increased number of interfaces and their myriad differences are expected to cause a significant amount of visual clutter and demand greater attention from pedestrians~\cite{hollander2020save, dey2022investigating, robert2019future, singer2020displaying,verstegen2021commdisk}. Information overload could easily lead to the inability to identify the most critical pieces of information~\cite{dey2022investigating, rossi2022impact} and, thus, pose a high level of safety risks~\cite{hollander2020save, hollander2019overtrust}. Traffic efficiency may also be adversely affected in multi-lane crossing scenarios where pedestrians must process multiple cues to cross the road safely~\cite{hollander2022take, wang2020adaptability}. \citet{hollander2022take} found that the individual AV signals (e.g., projected crosswalks) caused participants to stay longer in the first lane while waiting for the second lane to clear. Consequently, several studies have aimed to combine information from multiple vehicles and provide pedestrians with an augmented reality (AR) safety corridor~\cite{hesenius2018don, tran2022designing, tabone2021towards}.

\subsubsection{Multi-Lane Safety}

In a mixed traffic environment, pedestrians are likely to encounter AVs with varying degrees of automation alongside conventional vehicles. In light of this, \citet{mok2022stopping} and~\citet{moore2019case} mentioned the risks of eHMIs diverting pedestrian attention away from other road users and their motion cues. A VR-based study investigating pedestrian crossing behaviour in mixed traffic~\cite{mahadevan2019av}, however, revealed that participants acted more cautiously when encountering vehicles that did not communicate explicitly (i.e., those without an interface).

Multi-vehicle scenarios and mixed traffic conditions also necessitate caution in selecting the message perspective for eHMIs. Studies have found that egocentric messages suggesting a pedestrian action, such as `Walk', have higher clarity ratings and faster response times compared to allocentric messages communicating the AV's intent and states, such as `Braking'~\cite{eisma2021external}. As stated by \citet{eisma2021external}, the former perspective explains why AVs are stopping and does not require pedestrians to shift their mental perspective. Nonetheless, eHMI-design best practices are moving away from the use of advice and instructions due to concerns about safety and liability issues (e.g., ISO/TR 23049:2018~\cite{iso2018}). Firstly, instructional eHMIs may lessen pedestrians' responsibility in making crossing decisions, leading to an overreliance on the user interface~\cite{mahadevan2018communicating}. More importantly, such eHMIs risk creating a false expectation of the state of the surrounding (mixed) traffic, which is beyond the AV's control~\cite{andersson2017hello, habibovic2018communicating, dey2020taming, hollander2019investigating, de2022external, lagstrom2016avip}. For instance, a design concept where a zebra-crossing is projected onto the area in front of the vehicle could lead pedestrians to cross the subsequent lane when it might not be safe to do so~\cite{locken2019should, hollander2022take}. Given this, \citet{locken2019should} and \citet{colley2020evaluating} emphasised the importance of AVs considering the status of the other lanes when providing a crossing signal to pedestrians. One example involves the concept of the omniscient narrator, where a vehicle changes the communication message based on its knowledge of the traffic situation~\cite{colley2020towards}.

\subsubsection{Audibility}

\citet{mahadevan2018communicating} posited that multiple unsynchronised AVs communicating with pedestrians using audio messages may result in unpleasant noise rather than beneficial information. Investigating a traffic scenario with two AVs offering auditory cues, \citet{colley2020towards} found that some participants heard only one message or an echo when the vehicles arrived simultaneously. Furthermore, in densely populated urban areas, auditory cues may be drowned out by the noise arising from the high vehicle volume on the streets~\cite{mahadevan2019av}.

\subsubsection{Clarity of Recipients}

Directing eHMI communication messages at specific recipients is deemed imperative to ensure that pedestrians are confident they are being addressed by the AV, analogous to the acknowledgement conveyed by human eye contact. A study investigating different light-based eHMIs in a shared space setting revealed that most participants noticed the signal but did not assume it was intended for them, even though the scene involved only one pedestrian~\cite{hensch2019effects}. 

The directedness of eHMIs may be further compromised when multiple pedestrians are present, potentially leading to misunderstandings about which pedestrian the AV is attempting to communicate with~\cite{ackermans2020effects,colley2021investigating, dey2018interface,dey2020distance,tabone2021towards,ackermann2019experimental,benderius2017best,dey2021towards,dietrich2019projection,gui2022going,kaleefathullah2022external,kass2020standardized,prattico2021comparing,stromberg2018designing,wilbrink2021scaling,hollander2020save,eisma2021external,lagstrom2016avip,hollander2019investigating,de2022external,hollander2022take,robert2019future,dey2020taming,verstegen2021commdisk,hoggenmueller2022designing,tabone2021vulnerable,tran2022designing,li2018cross,singer2020displaying}. Studies have highlighted that the presence of other pedestrians introduces uncertainty to crossing scenarios~\cite{wilbrink2021scaling} and reduces the communication clarity of eHMIs~\cite{colley2021investigating}. Moreover, non-directed eHMIs significantly increased a pedestrian's willingness to cross compared to the baseline condition with no eHMI~\cite{wilbrink2021scaling, dey2021towards}, even in scenarios where the yielding message was intended for another~\cite{dey2021towards}. This finding suggests that without clear and directed communication, pedestrians are more likely to misinterpret AV communication and respond inappropriately, resulting in severe consequences in real-world traffic situations~\cite{dey2021towards}. Notably, the likelihood of undesirable or adverse outcomes may increase when eHMIs give commands or instruct pedestrians to cross~\cite{ackermann2019experimental,kass2020standardized,singer2020displaying,eisma2021external,lagstrom2016avip,hollander2019investigating,de2022external,robert2019future,tabone2021vulnerable}; a display that signals `Go ahead', for instance, could prompt non-targeted pedestrians to engage in risky behaviour. 

\subsubsection{Group Influence}

Research has indicated that pedestrians crossing in groups tend to be more careless and pay less attention to crosswalks and approaching traffic~\cite{rasouli2019autonomous}. It is, therefore, of value to determine whether such group influence remains when pedestrians interact with AVs equipped with eHMIs~\cite{lee2022learning}. \citet{mahadevan2019av} investigated the scenario of AI-based agents with different crossing behaviours (none, early crossers, and timely crossers) crossing the road alongside the participants in mixed traffic. Although the agents' behaviours did not significantly impact the participants' willingness to cross, half of the participants indicated that the agents may have affected their crossing strategy. In their observation study, \citet{chen2020autonomous} found stronger evidence of the prevalence of herd behaviour among pedestrians, with most individuals in a group crossing situation not making any effort to observe the traffic if another person was already present in the crosswalk. To investigate whether the effect of eHMIs is stronger than the influence of group pedestrian behaviour, \citet{colley2022effects} examined a situation where virtual pedestrians in close proximity to the participants ignored the AV's non-yielding intention. Although the authors were unable to determine which factor had a greater impact, they found that eHMIs continue to positively affect trust, cognitive load, and communication quality.

It should be noted that the actual behaviour of other pedestrians, rather than their mere presence, impacts participants' crossing behaviour~\cite{colley2022effects, colley2021investigating, joisten2021communication}. For instance, a standing pedestrian group that did not engage in any specific action (e.g., step onto the street to cross) had no effect on participants' crossing duration~\cite{colley2021investigating} and did not induce any imitable behaviour~\cite{joisten2021communication}. Nonetheless, participants in the study by \citet{joisten2021communication} felt that crossing the street in a group increased their safety and confidence as they believed the AV could more easily detect a group than a single pedestrian.

\subsubsection{Visibility}

An eHMI display located on the exterior of a vehicle or projected onto the road might present visibility issues, especially in shared spaces where pedestrians approach from various directions~\cite{ezzati2021interaction,risto2017human}. A co-design study conducted by \citet{asha2021co} also revealed that wheelchair users may encounter various visual obstructions, such as roadside elements and the presence of other pedestrians. A line of multiple vehicles one after the other further complicates the visibility of eHMIs. According to 3D simulation results~\cite{troel2019ehmi}, a frontal eHMI display is optimal only if a vehicle is the first one in a lane. This issue has led to the incorporation of multiple displays~\cite{macdonald2018}, a 360-degree LED light band~\cite{volvocars2018}, or a roof-mounted cylindrical interface~\cite{verstegen2021commdisk} in several eHMI concepts.

\subsubsection{Privacy}
In ride-sharing scenarios, directed eHMIs can help pedestrians identify the AV that is trying to pick them up~\cite{gui2022going}. However, eHMIs that disclose private information have received negative evaluations~\cite{lim2022ui}. Rather, when the goal is to easily locate a car in a crowd, using a distinctive colour or graphical symbol or comprising a combination of letters and numbers has been deemed acceptable~\cite{lim2022ui, volvocars2018, hoggenmueller2022designing}. Anthropomorphic robotic eyes that can convey fine-grained turning directions may also aid in identification~\cite{gui2022going}.

\subsection{General Remarks}

\subsubsection{Research Foci}

The \textit{Clarity of Recipients} (55.6\%, 30) has formed the primary concern of the research community, followed by \textit{Information Overload} (27.8\%, 15) and \textit{Multi-Lane Safety} (25.9\%, 14). This may be the case since the issue of recipient clarity extends beyond scenarios involving multiple pedestrians. In addition to pedestrians, other VRUs, such as cyclists~\cite{hollander2021taxonomy}, and drivers of conventional vehicles may also benefit from AV explicit communication. The inclusion of these diverse traffic participants may significantly increase the complexity of ensuring that AV communication reaches the intended road users. For instance, in examining the implicit and explicit communication of an AV interacting simultaneously with a pedestrian and a driver of a conventional vehicle, \citet{hubner2022external} highlighted an issue with recipient clarity, as a majority of the participants reported feeling wrongly addressed when the AV signalled the right of way to the other human road user.

Out of 54 selected papers, 16 (29.6\%) were empirical studies, suggesting that researchers are actively investigating eHMI scalability and seeking evidence to support their hypotheses and ideas. However, it also indicates that there may still be room for more empirical research to solidify the knowledge base in this area. Within this context, researchers have taken different approaches. Some have assessed established design concepts, such as the LED light bands, in more complex traffic scenarios~\cite{dey2021towards, wilbrink2021scaling, colley2022effects, rossi2022impact}. These studies contribute valuable insights regarding the underlying conditions, the severity of scalability issues, and whether they are significant enough to warrant further investigation. In contrast, other researchers have proposed new design concepts to address specific scalability issues, such as Smart Curbs~\cite{hollander2022take} and wearable AR solutions~\cite{hesenius2018don, tran2022designing}. It should be noted that at the time of the review, several new concepts had not undergone formal evaluation~\cite{tabone2021towards, verstegen2021commdisk} and were not counted as empirical studies. Nonetheless, the emergence of these ideas further emphasised the growing interest and innovation in scaling up AV external communication.

\subsubsection{Evaluation Method}

The identified empirical studies have been largely conducted in a controlled laboratory environment, with VR simulations being the most frequently used prototype representation (62.5\%). This finding suggests that VR is a suitable test bed for exploring the scalability of eHMIs as it allows for the design and evaluation of novel design concepts under various traffic scenarios while ensuring the safety of participants. According to a review of VR studies on AV–pedestrian interaction~\cite{tran2021review}, the effectiveness of VR simulations could be further enhanced through the use of more realistic AV driving behaviour and the inclusion of other road users in the same scenario using coupled/distributed simulators~\cite{bazilinskyy2020coupled}. The latter approach is particularly appealing as it is currently unclear how the use of human-controlled avatars rather than AI virtual agents might impact group influence and the clarity of recipients.

Notably, two of the studies evaluated eHMIs in a more realistic environment (i.e., a closed test track facility~\cite{rossi2022impact} and public roads~\cite{chen2020autonomous}). \citet{rossi2022impact} confirmed some problems previously identified in VR simulations~\cite{tran2022designing}, such as participants missing eHMI light patterns due to looking back and forth between the two AVs. Meanwhile, in their public experiment, \citet{chen2020autonomous} observed a stronger effect of group influence on pedestrians' crossing behaviour. In both studies, the use of physical prototypes helped elucidate how the visibility of the interface was affected by factors such as the sun or the angle at which it was viewed~\cite{rossi2022impact, chen2020autonomous}. These findings demonstrate the value of evaluating AV external communication under real-world traffic conditions.

\section{Discussion}

Despite the growing number of relevant empirical studies in recent years, the most pressing scalability issues of eHMIs have yet to be fully resolved. As a first step to structure future research efforts, we propose overarching research directions aimed at designing scalable AV-pedestrian interactions and reflect on the study limitations. 

\subsection{Research Directions based on Communication Locus}

According to the comprehensive review by \citet{dey2020taming}, around two-thirds of existing eHMI design concepts incorporated visual and auditory cues onto vehicles. This underlines the necessity for scalable vehicle-based eHMIs. Nevertheless, researchers have also proposed extravehicular solutions such as urban infrastructures and pedestrians' personal devices (e.g., wearables and smartphones)~\cite{dey2021towards, tabone2021vulnerable, dey2020taming}. As a result, three distinct communication loci, vehicle-based, infrastructure-based, and pedestrian device-based, are being explored simultaneously as potential solutions to address eHMI scalability issues. This section revisits the definition and characteristics of communication locus, aiming to establish a common understanding within the research community before delving into their potential in facilitating \textit{simplified} communication (presenting information in a clear, concise, and easily understandable manner), \textit{coordinated} communication (synchronising and harmonising the exchange of information between different systems), and \textit{targeted} communication (providing the right information to the right recipients at the right time). These three communication qualities are identified as particularly relevant in overcoming the most significant challenges related to eHMI scalability.

The locus of AV communication, or interface location, was originally defined by \citet{mahadevan2018communicating} in a participatory design study. This design dimension includes four categories based on where participants placed their interfaces: (1) vehicle-only, (2) vehicle and infrastructure, (3) vehicle and pedestrian, and (4) mixed. In this proposal, while the vehicle is always responsible for communicating with pedestrians, this responsibility may be shared with road infrastructures (e.g., traffic lights) and pedestrians' personal devices (e.g., smartphones). In the design space for AV external communication proposed by \citet{colley2020design}, the communication locus remains the same, though the categories are revised to represent independent placements, including those of the vehicle, infrastructure, and pedestrian. In this paper, we classified projection-based eHMIs (e.g., \cite{nguyen2019designing}) as part of the vehicle category, mainly because the projections originate from and reflect the vehicle's intentions even though they are not physically located on the vehicle.

The primary distinction between infrastructure and pedestrian-based eHMIs versus vehicle-based eHMIs is that the communication signals are detached from the vehicle. Being detached eHMIs, there exists two different roles infrastructures and personal devices can play:

\begin{itemize}
    \item As communication outlets: In this role, infrastructures and personal devices transmit communication signals from AVs to pedestrians. For instance, \citet{mahadevan2018communicating} used a smartphone to relay the message \textit{`I can see you'} to pedestrians~\cite{mahadevan2018communicating}. However, their findings showed that pedestrians considered this type of message more reliable when it originated from the vehicle rather than a third-party source, as it is closely connected to the vehicle's operation.    
    \item As communication sources: In this second role, infrastructures serve to facilitate traffic, such as the Starling Crossing, which adjusts its road markings in real time to guide pedestrians, cyclists, and drivers~\cite{umbrellium2017}. On the other hand, pedestrian devices offer personal guidance for navigating the environment; for example, a smartphone application might provide on-screen directions to enhance pedestrian situational awareness while crossing roads~\cite{hollander2020save}.
\end{itemize}

Distinguishing between these two roles is crucial because they may have diverse safety and liability implications, varied user experiences, and different technical requirements in terms of hardware, software, and communication protocols. Furthermore, the second role goes beyond the classic understanding of eHMIs, which mainly involve external interfaces that convey AV intentions and operational states to pedestrians and other VRUs. In the context of designing scalable AV-pedestrian interaction, our focus will be solely on the second role.

\subsubsection{Vehicle-based:} In densely populated urban areas, the presence of numerous vehicles each equipped with a unique eHMI can potentially increase the visual and auditory stimuli, contributing to an already dynamic and possibly overwhelming environment. Given the abundance of proposed eHMI design concepts, standardisation emerges as an essential and imminent direction to achieve simplified communication. Standardised cues not only make the learning process easier for road users but also reduce the variety of cues they encounter. This uniformity lessens the cognitive effort involved in interpreting and reacting to eHMI signals. The notion of less cognitive demand is based on the principle of cognitive economy,~\cite{colman2015dictionary} where our minds seek to minimise the amount of mental effort spent on processing stimuli. International Organisation for Standardisation (ISO) has established a working group\footnote{\url{https://www.iso.org/committee/5391225.html}, last accessed April 2023} (ISO/TC 22/SC 39/WG 8) dedicated to advanced driver assistance systems, automated driving, and eHMI standardisation for external communication. The process of standardising eHMIs is intricate and ongoing, requiring close collaboration among diverse stakeholders in the automotive industry. These stakeholders must determine which dimensions to standardise while considering individual, cultural, and regulatory differences to ensure the effectiveness and widespread adoption of the resulting standards.

One important research direction in the development of eHMIs is determining under which circumstances eHMIs are truly necessary, as opposed to maintaining an always-on mode of operation. This approach may help to prevent excessive stimuli and conserve system resources such as energy and processing power. In line with this, \citet{dey2022investigating} examined the need for an eHMI to communicate an AV’s non-yielding intent. However, the findings indicated a preference for eHMIs that communicate an AV’s intent explicitly at all times, highlighting the need to determine a threshold beyond which the information provided becomes detrimental~\cite{mahadevan2018communicating}.

Another solution involves using interconnected eHMIs, where a single vehicle communicates the intentions of all other vehicles in the vicinity~\cite{colley2020towards, hollander2022take}. Such eHMIs, enabled by connected vehicle technology, can potentially lower pedestrians' cognitive load by minimising the number of communicating AVs and enhance pedestrians' safety in heavy traffic by tailoring the communication to the prevailing traffic conditions. However, preliminary findings have been mixed due to pedestrians' unfamiliarity with the concept of connected vehicles~\cite{colley2020towards}. From a technical perspective, the interconnected eHMI approach adds complexity to communication protocols, such as deciding which AV takes on the primary communication role. Addressing potential challenges related to data security and issues like delays or loss of connectivity is also crucial to maintain pedestrian safety. As a result, further research is necessary to determine the viability of interconnected eHMIs as a design concept, taking into account both human factors and implementation feasibility.

With respect to targeted communication, vehicle-based eHMIs can direct their messages in a specific direction. For example, \citet{gui2022going} investigated how robotic eye gazes can convey an AV's fine-grained moving directions, enabling pedestrians to clear the path in an open space or determine if the AV (taxi) intends to pick them up. The Volvo 360c concept incorporated directional warning sounds aimed at pedestrians who unexpectedly step onto the road~\cite{volvocars2018}. \citet{kaup2019interact} proposed the concept of directed signal lamps capable of addressing multiple road users separately with discrete light channels. Another approach to achieving targeted communication involves presenting a pedestrian's position on an eHMI display in a simplified schematic form (e.g., a dot~\cite{verstegen2021commdisk} or a strip~\cite{nissan2015}) or adjusting communication based on the distance between the vehicle and the pedestrian (distance-dependent eHMIs)~\cite{dey2020taming, dey2020distance}. According to the taxonomy established by~\citet{dey2020taming}, all these design concepts can be classified as partially scalable, meaning they can address multiple road users concurrently but only to a certain extent. We argue that this limited degree of scalability may not necessarily be viewed a drawback, given that most traffic situations and interactions usually involve a manageable number of road users. Instead, the more critical aspect is the eHMI's ability to deliver unambiguous information about whether it is safe for specific road users to proceed. For example, road projections visualising the precise location at which the vehicle will stop can clearly convey the intended recipient~\cite{dey2022investigating}. As a result, high communication resolution should be made a key design goal for eHMIs and must be thoroughly assessed in scenarios that are likely to occur, such as urban intersections and pick-up/drop-off zones.

 \subsubsection{Infrastructure-based:} The majority of infrastructure-based design concepts leverage existing traffic elements that are both familiar to pedestrians and universally understood, such as zebra crossings~\cite{umbrellium2017, locken2019investigating, hoggenmueller2019enhancing}. By integrating interactive features into these well-known elements, infrastructure-based designs may enable pedestrians to quickly adapt to the new systems without facing a steep learning curve. In situations with multiple vehicles, infrastructure-based systems provide a centralised source of information for pedestrians, making it easier to process the data without having to attend to multiple vehicle-based displays simultaneously. 

One primary advantage of infrastructure-based concepts in coordinated and targeted communication lies in their integration of adaptive traffic management systems. These systems can dynamically adjust traffic control measures by leveraging real-time traffic data gathered from various sources, such as sensors, cameras, and connected networks. For instance, the Starling Crossing~\cite{umbrellium2017} adapts its configuration to different road users, environmental conditions, times of day, traffic volumes, and pedestrian behaviours, including distractions or haste. Similarly, Smart Curbs~\cite{hollander2022take} extend the traditional use of in-ground LED traffic signal lights~\cite{korea2021}. They detect approaching cars and use LEDs embedded in the curbs to communicate crossing instructions to pedestrians based on their current position. These systems can prioritise pedestrian crossings, promoting city walkability while optimising traffic flow~\cite{hollander2022take}. Additionally, infrastructure-based concepts are often perceived by road users as having more authority in facilitating traffic~\cite{tran2022designing}, which may contribute to a higher level of trust and compliance.

Despite the potential of smart infrastructures to enhance urban environments and transportation systems, the research community also acknowledges the challenges it presents, such as high initial development costs, ongoing maintenance, and regular upgrades, which can be particularly burdensome for resource-limited cities and developing countries~\cite{tabone2021vulnerable}. Other concerns include the impact of signal interference and network congestion on the reliability of wireless communication~\cite{tabone2021vulnerable}, as well as the potential for vandalism or tampering to cause system malfunctions or inaccurate data~\cite{colley2020towards}. Consequently, future directions should focus on developing cost-effective, resilient solutions and strategically implementing them in areas where they can significantly improve safety and efficiency, such as pedestrian-dense environments and accident-prone zones. Alongside these efforts, it is crucial to emphasise user-centred design and human factors research to prevent overreliance on technology and ensure individuals maintain a sense of personal responsibility while navigating traffic.

\subsubsection{Pedestrian-based:} Leveraging wearables~\cite{hesenius2018don, tran2022designing} and smartphones~\cite{hollander2020save, fitzgerald2020}, these design concepts offer the potential to consolidate communication cues from multiple AVs and enable communication with numerous relevant pedestrians~\cite{dey2020taming}. This approach enables tailored communication based on individual preferences and needs and supports context-aware communication by accounting for factors such as the user's location, speed, and direction of movement. Furthermore, the advent of connected vehicle technology may allow personal devices to function as virtual pedestrian light systems, creating an innovative, infrastructure-free means of requesting pedestrian right-of-way at intersections~\cite{tonguz2021system}.

While Vehicle-to-Pedestrian (V2P) smartphone applications continue to be the primary focus in research for AV-pedestrian communication~\cite{sewalkar2019vehicle}, there is a growing interest in wearable AR technology as an emerging research avenue~\cite{tabone2021vulnerable, tran2022designing, tabone2021towards, hesenius2018don, prattico2021comparing, tong2021augmented}. This interest stems from the potential advantages of wearable AR devices over traditional smartphone applications, such as offering a more immersive and intuitive experience, enhancing situational awareness, and enabling hands-free operation~\cite{starner1997augmented}. Furthermore, wearable AR can bolster spatial awareness by delivering contextual information about the user's environment, including the location and trajectory of nearby vehicles, potential hazards~\cite{tong2021augmented}, or suggested routes~\cite{hesenius2018don}. However, wearable AR devices may face challenges in terms of long-term use or user acceptance due to their form factor, weight, and design~\cite{billinghurst2021grand, azuma2019road}. Additionally, these devices, by design, can collect more personal data like location and gaze direction compared to smartphones, potentially raising privacy concerns for some users. More research is needed to address these challenges and develop user-friendly, privacy-preserving solutions that can facilitate widespread adoption. However, in order to foster inclusivity and equity in AV-pedestrian communication, it is crucial not to rely solely on wearable AR technologies~\cite{tabone2021vulnerable}. 

\renewcommand{\arraystretch}{1.8}
\begin{table*}[t]
\footnotesize
\centering
\caption{Overview of potential research directions for enhancing scalability in AV  external communication: A comparative analysis based on Vehicle, Infrastructure, and Pedestrian loci.}
\begin{tabular}{@{\hspace{5mm}}p{2cm}@{\hspace{1mm}}@{\hspace{5mm}}p{2.5cm}@{\hspace{5mm}}@{\hspace{5mm}}p{3cm}@{\hspace{5mm}}@{\hspace{5mm}}p{3cm}@{\hspace{5mm}}}

\textbf{} & \raisebox{-\totalheight}{\includegraphics[width=2.5cm]{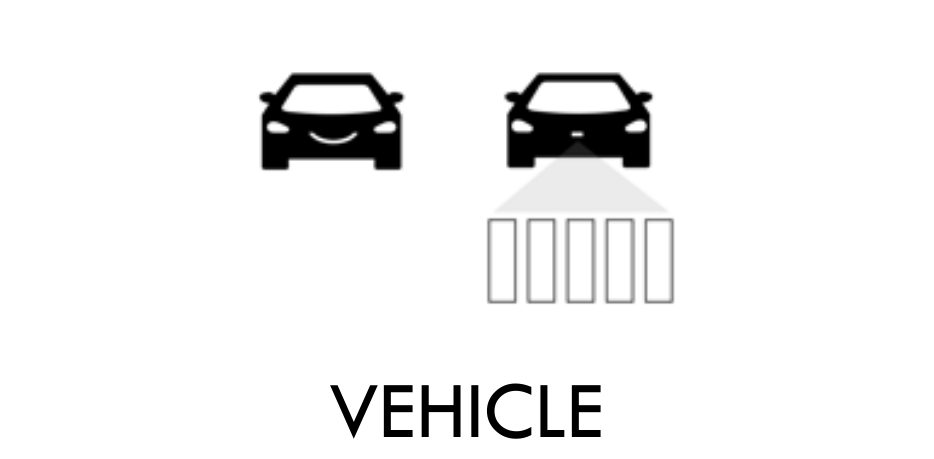}} \newline & \raisebox{-\totalheight}{\includegraphics[width=2.5cm]{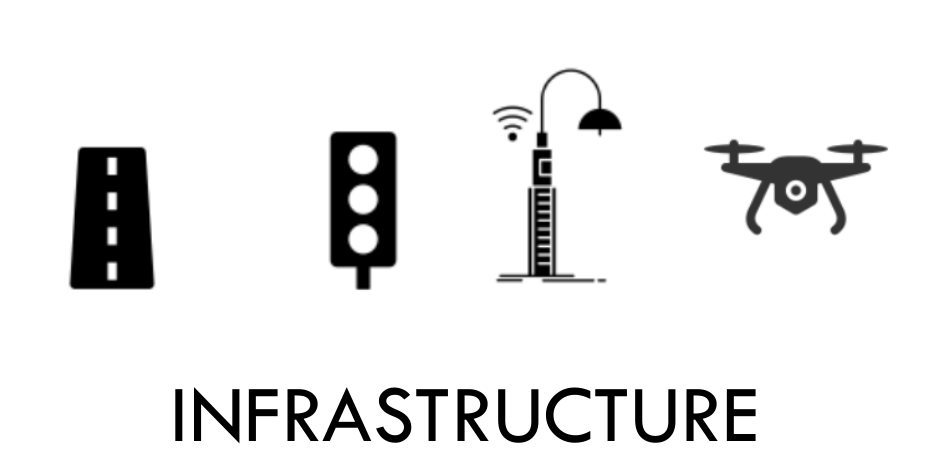}} \newline  & \raisebox{-\totalheight}{\includegraphics[width=2.5cm]{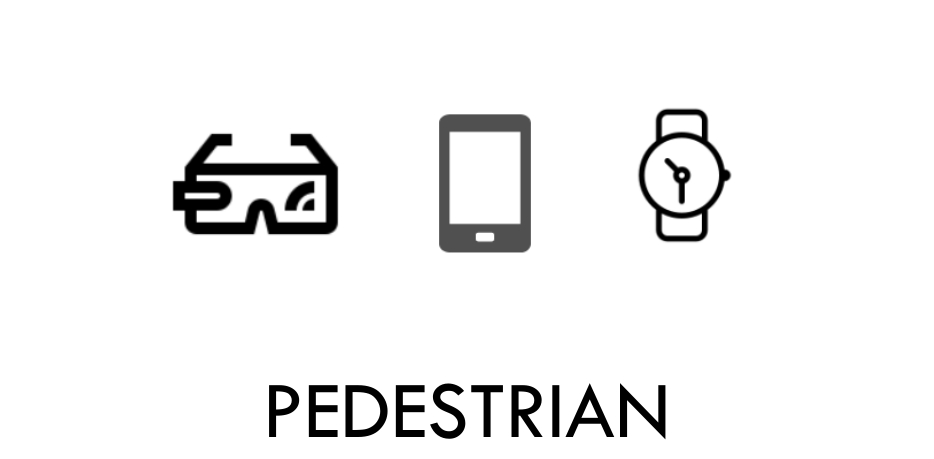}} \newline  \\ 
\hline
\textbf{Simplified} & Standardised cues \newline On-demand operation\newline Interconnected eHMIs & Central source of information & Cue consolidation\\ 
\hline
\textbf{Coordinated} & Interconnected eHMIs & Adaptive traffic management  & V2P applications \\ 
\hline
\textbf{Targeted} & High-resolution \newline communication & Instructions based on \newline pedestrian position & Tailored communication \newline via wearables/smartphones \\ 
\hline
\end{tabular}
\label{table:loci_summary}
\end{table*}

\subsubsection{Summary} A comparative analysis of three communication loci—vehicle, infrastructure, and pedestrian based on their potential for simplified, coordinated, and targeted communication is summarised in \autoref{table:loci_summary}. In the pursuit of designing scalable AV-pedestrian interactions, each of these loci exhibits distinct benefits and drawbacks. However, most of them are derived from theoretical reasoning and assumptions rather than empirical data. This necessitates further research and validation through real-world experiments, user studies, and data-driven approaches. Additionally, exploring multiple communication loci rather than focusing on just one enables the development of more adaptable, reliable, and inclusive eHMI solutions that cater to a wide range of situations, user needs, and abilities. By considering the strengths and weaknesses of each communication locus, researchers can identify potential synergies and develop more effective, evidence-based designs that enhance the safety and efficiency of AV-pedestrian interactions in diverse contexts. In fact, a hybrid loci approach, which incorporates built-in redundancy and increased reliability, has been previously explored and recommended in the works of \citet{mahadevan2018communicating} and \citet{locken2019investigating}, demonstrating the potential value of combining multiple communication loci in eHMI solutions.

\subsection{Limitations}

Our work has several limitations. Primarily, the comprehensiveness of our review could be impacted by our choice of keywords and selection strategy, potentially resulting in the omission of relevant studies. Moreover, the lack of multiple coders independently analysing the data could raise questions about the consistency of data interpretation and limit the generalisability of our findings. Within the scope of this paper, our primary emphasis is on AV-pedestrian interactions. We made a conscious decision to not delve into the complex dynamics involving other road users such as cyclists, motorbike riders, and conventional drivers. This focus, whilst necessary for the initial exploration, might lead to an incomplete understanding of the overall communication ecosystem and limit our insights on the scalability of eHMIs across different types of road users.

\section{Conclusion}

Our review revealed that since the 2020 study by \citet{colley2020unveiling}, which reported eHMIs being predominantly evaluated in simple traffic scenarios, there has been a growing focus within the automotive research community on the ability of eHMIs to handle larger numbers of pedestrians and vehicles in complex traffic environments. Analysing 54 papers discussing scalability issues, including 16 empirical studies examining eHMIs in multi-vehicle or multi-pedestrian situations, allowed us to identify and organise the challenges faced by AV external communication. Our review offers researchers insight into the possible challenges of implementing eHMIs in real-world situations. Additionally, we have highlighted essential communication qualities required for effective eHMIs in complex situations and proposed high-level research directions focused on three communication loci: vehicle, infrastructure, and pedestrian, to guide future efforts in this domain.

\begin{acks}
This research is supported by an Australian Government Research Training Program (RTP) Scholarship and through the ARC Discovery Project DP200102604, Trust and Safety in Autonomous Mobility Systems: A Human-centred Approach. We extend our gratitude to the anonymous reviewers for their insightful comments and suggestions. Additionally, we thank the coordinator for their invaluable assistance throughout the shepherding process. Their efforts have significantly contributed to the final version of this paper.
\end{acks}

\bibliographystyle{ACM-Reference-Format}
\bibliography{references}

\end{document}